\documentclass[%
 aip,
 amsmath,amssymb,
 reprint,%
]{revtex4-1}

\usepackage{graphicx}% Include figure files
\usepackage{dcolumn}% Align table columns on decimal point
\usepackage{bm}% bold math
\usepackage[caption=false]{subfig}

\usepackage[utf8]{inputenc}
\usepackage[T1]{fontenc}
\usepackage{mathptmx}
\usepackage{float}
\usepackage[british]{babel}
\usepackage{makecell}

\begin{document}

\preprint{AIP/123-QED}

\title[Locality, physical invariants, and falling liquid film control]{Exploiting locality and physical invariants to design effective Deep Reinforcement Learning control of the unstable falling liquid film \\~\\}
% Force line breaks with \\

\author{Vincent Belus} \email{vincent.belus@mines-paristech.fr}
\affiliation{ 
MINES Paristech PSL, Research University CEMEF, 06904 Sofia Antipolis Cedex, France \\
}
\affiliation{ 
Department of Mathematics, University of Oslo, 0316 Oslo, Norway \\
}%

\author{Jean Rabault} \email{jean.rblt@gmail.com, \textbf{corresponding author}}
\affiliation{ 
Department of Mathematics, University of Oslo, 0316 Oslo, Norway \\
}%
\affiliation{ 
MINES Paristech PSL, Research University CEMEF, 06904 Sofia Antipolis Cedex, France \\
}%

\author{\\Jonathan Viquerat} \email{jonathan.viquerat@mines-paristech.fr}
\affiliation{ 
MINES Paristech PSL, Research University CEMEF, 06904 Sofia Antipolis Cedex, France \\
}%

\author{Zhizhao Che} \email{chezhizhao@tju.edu.cn}
\affiliation{ 
State Key Laboratory of Engines, Tianjin University, Tianjin 300072, China \\
}%

\author{Elie Hachem} \email{elie.hachem@mines-paristech.fr}
\affiliation{ 
MINES Paristech PSL, Research University CEMEF, 06904 Sofia Antipolis Cedex, France \\
}%

\author{Ulysse Reglade} \email{ulysse.reglade@mines-paristech.fr}
\affiliation{ 
MINES Paristech PSL, Research University CEMEF, 06904 Sofia Antipolis Cedex, France \\
}%
\affiliation{ 
Department of Mathematics, University of Oslo, 0316 Oslo, Norway \\
}%

\date{\today}% It is always \today, today,

\begin{abstract}
    Instabilities arise in a number of flow configurations. One such
    manifestation is the development of interfacial waves in multiphase flows,
    such as those observed in the falling liquid film problem. Controlling the
    development of such instabilities is a problem of both academic and
    industrial interest. However, this has proven challenging in most cases due
    to the strong nonlinearity and high dimensionality of the underlying
    equations. In the present work, we successfully apply Deep Reinforcement
    Learning (DRL) for the control of the one-dimensional (1D) depth-integrated
    falling liquid film. In addition, we introduce for the first time
    translational invariance in the architecture of the DRL agent, and we
    exploit locality of the control problem to define a dense reward function.
    This allows to both speed up learning considerably, and to easily control
    an arbitrary large number of jets and overcome the curse of dimensionality
    on the control output size that would take place using a naive approach.
    This illustrates the importance of the architecture of the agent for
    successful DRL control, and we believe this will be an important element in
    the effective application of DRL to large two-dimensional (2D) or
    three-dimensional (3D) systems featuring translational, axisymmetric or other
    invariants.
\end{abstract}

\maketitle

\section{Introduction}

Falling liquid films are a common phenomenon both in industry and nature
\cite{doi:10.1002/aic.690310907, nosoko1996characteristics,
doi:10.1063/1.2158428, doi:10.1063/1.2750384}. Such flows are highly complex
due to their nonlinearity, and the presence of an interface between the liquid
and gas phases. In addition, there are many instabilities taking place in such
flows as highlighted by the previous references. These are both a challenge and
an attraction for engineers and scientists. Progressing towards effective
strategies for the control of instabilities in falling liquid films, is
therefore a relevant and interesting problem. Some work has been performed in
the case of falling liquid flows \cite{doi:10.1063/1.3294884,
doi:10.1063/1.4938761, doi:10.1063/1.2196450}, but the design of general,
robust control methods that can be adapted to specific applications in a
flexible way without user expertise is still a relevant problem. Finding such
general control laws is made complex due to the combination of strong
non-linearity, high dimensionality, and time-dependence of those systems.

However, in recent years, methods based on data-driven approaches inspired by
recent results from the Machine Learning community have proven increasingly
successful. Those include several classes of methods, such as Genetic
Programming (GP) \cite{banzhaf1998genetic, langdon2013foundations}, and Deep
Reinforcement Learning (DRL) \cite{sutton1998introduction, mnih2015human}. These
methods are now being applied to Fluid Mechanics, with a series of recent
successes that include, for example, controlling complex wake dynamics in
two-dimensional (2D) simulations
\cite{rabault_kuchta_jensen_reglade_cerardi_2019, doi:10.1063/1.5116415},
controlling chaotic model Partial Differential Equations (PDE) systems
\cite{duriez2017machine, bucci2019control} , and a number of drag and vortex
shedding control strategies \cite{gautier2015closed, doi:10.1063/1.5115258,
doi:10.1063/1.5055016}. However, one must be able to scale up those methods,
both in terms of number of simulations and number of control outputs, in order
to envision control of realistic situations. While the first scaling problem
has recently been tackled and proven to work well \cite{doi:10.1063/1.5116415},
demonstrating the ability of such methods to handle well a large number of
outputs without hitting the curse of dimensionality remains a critical open
problem.

In the present work, we consider the 1D falling liquid film problem and its
optimal control through a DRL approach using small localized actuators. This
problem is well suited for exploring optimal control of systems with many
actuators, as it is both strongly nonlinear, featuring the development of large
unstable interfacial waves, as well as inexpensive and quick to solve.
Therefore, it is an excellent model problem to explore the potential of DRL
applied to systems with many control signals, as it allows fast prototyping,
training, and assessment of different methodologies. Our contribution in this
article is double: first, we show that this system can be very efficiently
controlled using DRL. Second, we discuss different variations around how DRL can be
applied in practice to such a problem with a potentially large control space
dimensionality. There, we show that different approaches are possible to take
advantage of the invariance by translation of the system, and that the choice
of the method used has a large impact on the quality of the control strategy
obtained as well as on the speed of learning. In addition, we observe that the
locality of the system allows to define a dense reward function, which provides
a fine-grained training signal and there also allows both better and faster
training.

The organization of this manuscript is as follows. First, we present the
methodology used for both the 1D falling liquid film simulation, and the DRL
methodology including the different strategies to effectively implement
multiple-output control. Then, we present the results obtained controlling the
system, and we compare the efficiency of these different strategies. Finally,
we discuss the applicability of our findings to different control problems,
both within Fluid Mechanics and at large.

\section{Methodology}

\subsection{Falling liquid film simulation}

In this work, we consider a liquid film that flows down an inclined plane. The
$x$ coordinate is chosen along the streamwise direction, i.e.\ following the
inclined plane. The formulation of the problem and the numerical scheme
implemented to solve it are similar to that in \cite{che2014ensemble}. More
specifically, the liquid chosen is an incompressible, Newtonian fluid with
constant properties. Those are its surface tension $\sigma$, viscosity $\mu$,
and density $\rho$. As a model for the falling film, we use the dimensionless,
depth-integrated system \cite{CrasterMatar2009RMP}:

\begin{equation}
    \frac{\partial h}{\partial t} + \frac{\partial q}{\partial x} = 0, \\
\end{equation}

\begin{equation}
    \frac{\partial q}{\partial t} + \frac{6}{5} \frac{\partial}{\partial x} \left( \frac{q^ 2}{h} \right) = \frac{1}{5 \delta} \left( h \frac{\partial^3 h}{\partial x^3} + h - \frac{q}{h^2} \right),
\end{equation}

\noindent where $h$ is the non-dimensional local film thickness, $q$ the
non-dimensional local flow rate, $\delta = (\rho H_c^{11}g^4/ \sigma)^{1/3} /
45 \nu^2$, with $H_c$ the film thickness without waves, $g = 9.81$ m/s$^2$ the
acceleration of gravity, and $\nu = \mu / \rho$. In practise, we will use
$\delta = 0.1$ in the following, similar to \cite{che2014ensemble}. This
formulation resorts on a semiparabolic velocity profile and satisfies the
no-slip boundary condition at the wall, as well as the zero stress boundary
condition at the gas-liquid interface. The boundary conditions at the inlet and
outlet are:

\begin{equation}
    h = 1,~ q = 1 \text{~ at~} x = 0,
\end{equation}

\begin{equation}
    \frac{\partial h}{\partial x} = 0,~ \frac{\partial q}{\partial x} = 0 \text{~ at~} x=L,
\end{equation}

\noindent where $L=300$ is the length of the domain. 
This value of $L$ is long enough for the development of different types of
waves to take place in the case without control. The initial condition in time
is obtained by simulating a uniform liquid film of thickness and mass flow rate
unity ($h=1$ and $q=1$) until the waves are fully developed.

Similarly to \cite{che2014ensemble}, the equations (1) and (2) are discretized
using the finite difference method. The transient terms are integrated using
the third order Runge-Kutta method (RK-3) \cite{Osher2003LevelSetBook}.
Convective terms are discretized using the Total Variation Diminishing (TVD)
scheme \cite{Versteeg2007FVMbook}. The grid size is $\Delta x = 0.1$ and the
time step is $\Delta t = 0.001$. In addition, we use a similar technique to
\cite{che2014ensemble} and add noise on the $h$ variable at the inlet of the
domain ($x=0$) to trigger the appearance of waves. This is done by replacing
(3) with:

\begin{equation}
    h(t) = 1 + r(t) \text{~ at~} x=0,
\end{equation}

\noindent where $r(t)$ is random, uniformly distributed in $[-5 \times 10^{-4};
5 \times 10^{-4}]$. In their work, \cite{che2014ensemble} have studied the
influence of the white noise input and found that its amplitude and
distribution does not have a significant effect on the overall behavior of the
waves due to the amplifier nature of the flow at specific frequencies.

In addition, we introduce forcing terms in the equations at several
user-tunable positions. In the following, we will refer to these individual
forcings as `jets'. The strength of the jets is set by the DRL algorithm (see
next paragraph) when applying control on the system. For simplicity, the
forcing is performed on the mass flow rate $q$, by adding the following
parabolic profile suction/blowing forcing $\delta q_i$ at each time step in the
numerical solver:

\begin{equation}
    \delta q_i(x, t) =
    \begin{cases}
        A_i(t) \cdot (x - l_i) (r_i -x) & \text{if }\ l_i < x < r_i,\\
        0 & \text{otherwise},
    \end{cases}
\end{equation}

\noindent where $i$ (integer between 1 and $N$, the total number of jets) is
the index of the jet currently considered, which is located between
$x$-positions $l_i < x < r_i$, and $A_i(t)$ is the strength of the
corresponding jet at time $t$. As visible in (6), this corresponds to using a
small jet following a parabolic profile, going to zero on the right and left
edges of each of the forcing areas, with the centers being located at positions
$c_i = (l_i + r_i) / 2$ and the jets having half-widths $w_i = (r_i - l_i) /
2$. In the following, the maximum strength of the jets, as well as their widths
and locations, will be used as physical meta-parameters of the flow
configuration. Note that both injection of mass (positive forcing corresponding
to an increase of the local mass flow rate, i.e. blowing) and removal of mass
(negative forcing, corresponding to a reduction in the local mass flow rate,
i.e. suction) are possible.

\begin{figure*}
\begin{center}
\includegraphics[width=.95\textwidth]{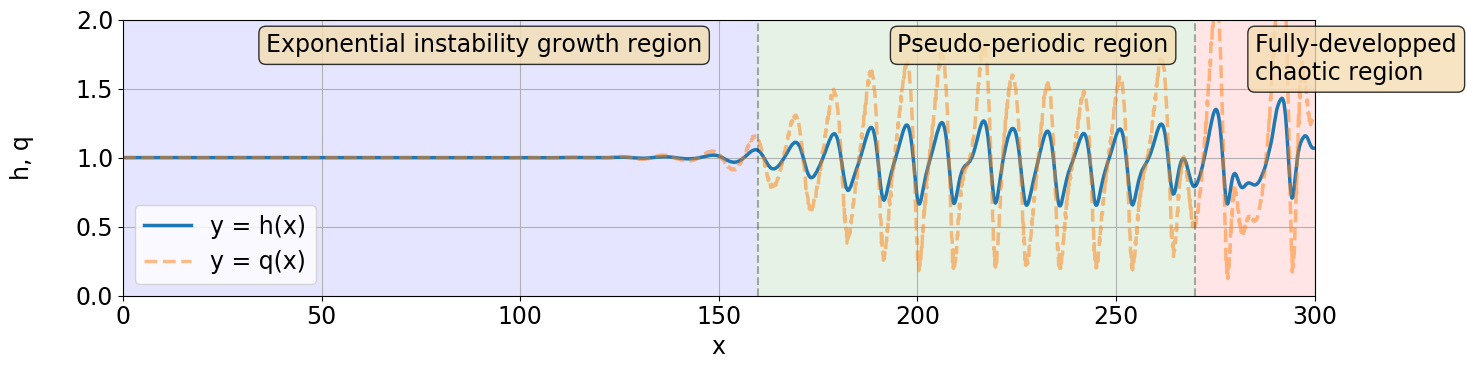}
\caption{\label{fig:simu} Illustration of a converged falling liquid film simulation
    performed with zero control but with inlet disturbances.
    Three areas are clearly visible: first, a region where the flow disturbances
    induced by the inlet boundary condition perturbations grow exponentially. Second,
    a region where the waves are pseudo periodic and get unstable. Third, a region
    where fully-developed chaotic behavior of the waves is observed.}
\end{center}
\end{figure*}

Those numerics are implemented in highly tuned C++ code for optimizing the
speed of execution, and made available to the high-level Python DRL library
(see next paragraph) through the use of C++/Boost Python bindings. All the
implementation is made available as Open Source (see the Appendix A). Using our
implementation, a simulation covering non-dimensional times $t=0$ to $t=200$
typically takes less than $30$ seconds on a modern CPU using a single core. The
problem is small enough that a large part of it can reside purely in CPU cache,
which also greatly improves performance. Typical converged simulation results,
with the inlet perturbation but without jet control, are illustrated in Fig.
\ref{fig:simu}.

In order to provide the input (or state observation) and reward to the DRL
agent, we use small regions in the neighborhood of each jet. The state is
obtained by reading from the simulation both $h$ and $q$ and considering them
as two different input channels. In all the following, unless stated otherwise,
both $h$ and $q$ are sampled in an area $A_{obs,i} = [r_i - L_{obs}; r_i]$,
where $L_{obs}$ is the size of the observation area. Similarly, the reward is
computed either locally on the right of each jet based on an area $A_{reward,i}
= [l_i; l_i + L_{reward}]$, or globally on the union of these areas
$\bigcup\limits_{i=1}^{N} A_{reward,i}$. Typical values are $L_{obs} = 25$ and
$L_{reward} = 10$ in the following.

The formula for the reward is: 

\begin{equation}
    \mathrm{R}(A_{reward}, t) = 1 - \chi \cdot \sqrt{\frac{\sum\limits_{x \in A_{reward}} [\mathrm{h}(x, t)-1]^2}{\mathrm{card}(A_{reward})}},
    \label{eqn:reward}
\end{equation}

\noindent where $\chi$ is a parameter chosen so that the reward calculated on
an environment without any control is close to 0, typically $\chi~\approx~5.7$.
$\mathrm{card}(A)$ is the number of elements in the set A. $A_{reward}$ is the
domain where we compute our reward, it can either be $A_{reward,i}$ or
$\bigcup\limits_{i=1}^{N} A_{reward,i}$ depending on what method we use (see
next section).

\begin{figure*}
    \begin{center}
    \includegraphics[width=.95\textwidth]{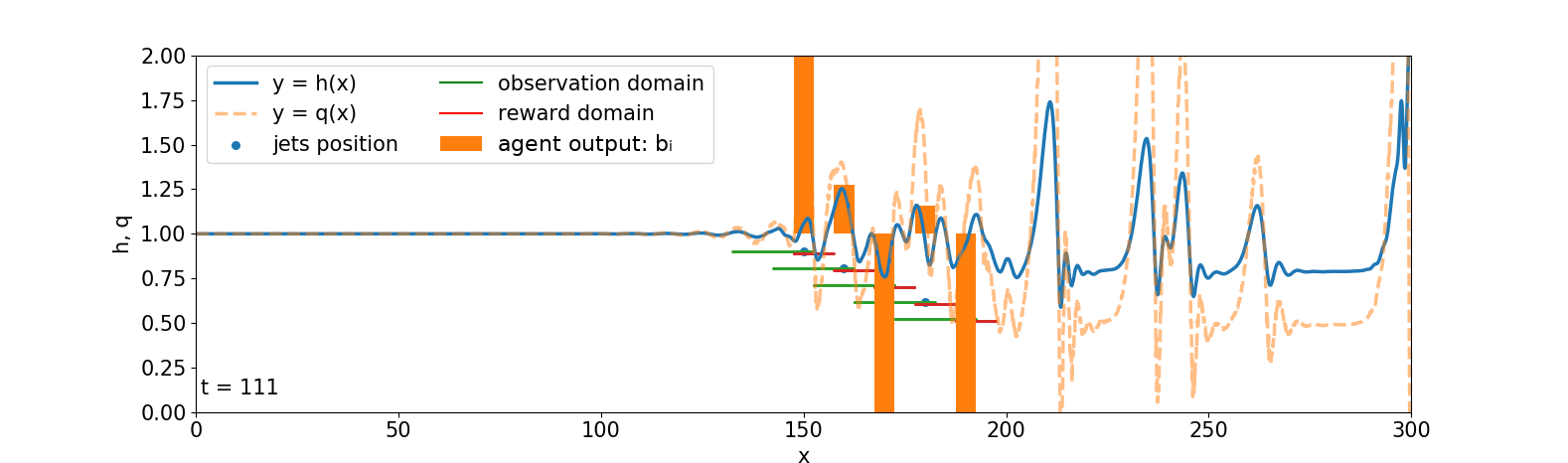}
        \caption{\label{fig:state_reward_config}
        Illustration of the observation space, reward
        space, and jet position during training, here with 5 jets.
        The forcing by the agent is illustrated by plotting directly the output of the ANN,
        which is between -1 and 1 and used to compute the control following
        Eqn. (\ref{scaling_eqn}). For clarity of the figure, the ANN
        output $b_i$ is shifted by an offset of +1 relatively to the vertical axis
        used for $h$ and $q$.}
        \end{center}
\end{figure*}

Using this reward, the network gets an incitation towards killing waves, and a perfect
reward is obtained when no waves at all are present ($h=1$ uniformly on the whole
reward domain), while any fluctuations in $h$ get penalized. 

$L_{obs}$ and $L_{reward}$ are to be chosen carefully. The reward being a single
value, it is essential for it to encapsulate relevant information about how our
action impacted the environment. We can suspect that a too large reward space
makes the reward less relevant about the effect of our actuation, while a too small
reward space may have difficulties capturing the effect of the control behind of the jet. 

In addition to this definition of the state and actions, a renormalization is
applied before the data are fed to the agent. The aim of this renormalization is
to make sure that the resting value of our data is 0 instead of 1, and that it
does not exceed a certain threshold in absolute value (typically, the maximum
output to the ANN should be approximately between 1 and 10), which is a
necessary condition for the DRL control to perform well. This renormalization is
performed by defining the state effectively given to the ANN as:

\begin{equation}
    \mathrm{h_{norm}}(A_{obs}, t) \cup \mathrm{q_{norm}}(A_{obs}, t),
\end{equation}

\noindent where

\begin{equation}
    \mathrm{h_{norm}}(x, t) = \mathrm{clip}(\gamma_h \cdot [\mathrm{h}(x, t) - 1], -S_{max}, S_{max}),
\end{equation}    

\noindent and

\begin{equation}
    \mathrm{q_{norm}}(x, t) = \mathrm{clip}(\gamma_q \cdot [\mathrm{q}(x, t) - 1], -S_{max}, S_{max}),
\end{equation}

\noindent where $\gamma_h \approx 1.0$ and $\gamma_q \approx 1.0$ are
normalization parameters, and $S_{max} \approx 5.0$ is the maximum value we are
ready to feed our ANN. Similarly the action effectively applied on the
simulation is:

\begin{equation}
    \mathrm{A_i}(t) = \frac{M \cdot b_i}{w_i^2}
    \label{scaling_eqn}
\end{equation}

\noindent where $b_i$ is the action effectively produced by the ANN, which is in
the range [-1, 1], and M is a hyperparameter defining the maximum strength of the
jets, typically
$M = 5$. $w_i$ is the half-width of the jets, as previously defined.

A typical illustration of the positioning of jets, as well as
the associated state and reward areas, is presented in Fig.
\ref{fig:state_reward_config}. In Fig. \ref{fig:state_reward_config}, as in similar figures in the
following of the paper, we present snapshots of the state of the system ($q$,
$h$) together with snapshots of the outputs $b_i$, $i=1..N$, provided by the
ANN. Those outputs are between 1 and -1, and displayed shifted by an offset of
+1 relatively to the vertical axis for clarity. Observe that the control effectively
applied is obtained by applying scaling proportional to $M$, as indicated in
Eqn. (\ref{scaling_eqn}).

In all the following, trainings are always started from a well-converged state
of the system, with fully developed waves being present. This corresponds to an
initial configuration of the system similar to what is visible in Fig. \ref{fig:simu}.
The maximum jet intensity is large enough that bad choices of the instantaneous
strength of the jets can create numerical blowup of the simulation. In this case,
the simulation is terminated, a negative reward of -5 is provided to
the ANN to `punish' it, and the simulation is resetted to the initial converged
state before training is resumed.

\subsection{DRL algorithm and strategies for multiple controls}

\begin{figure*}
\begin{center}
\includegraphics[width=.99\textwidth]{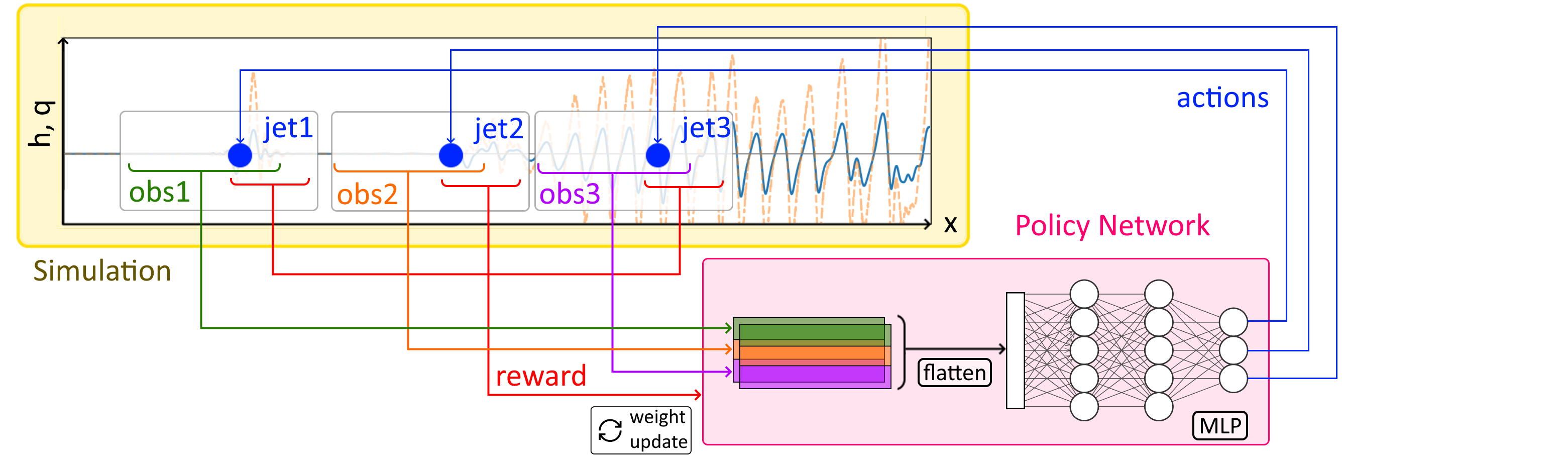}\\
    \vspace{0.5cm}
\includegraphics[width=.99\textwidth]{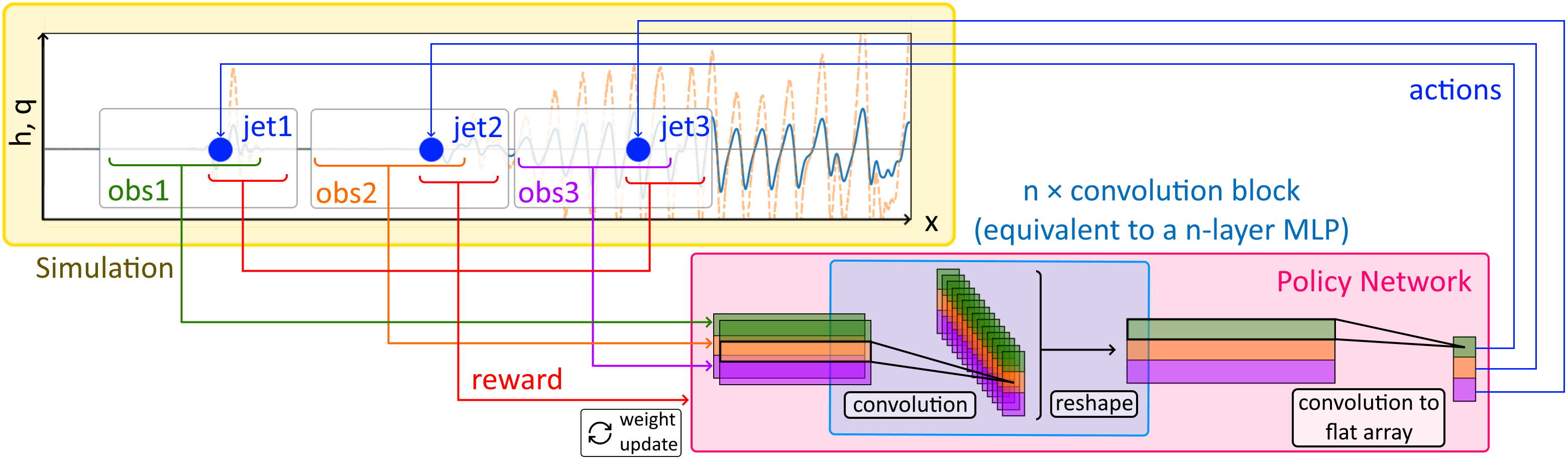}\\
    \vspace{0.5cm}
\includegraphics[width=.99\textwidth]{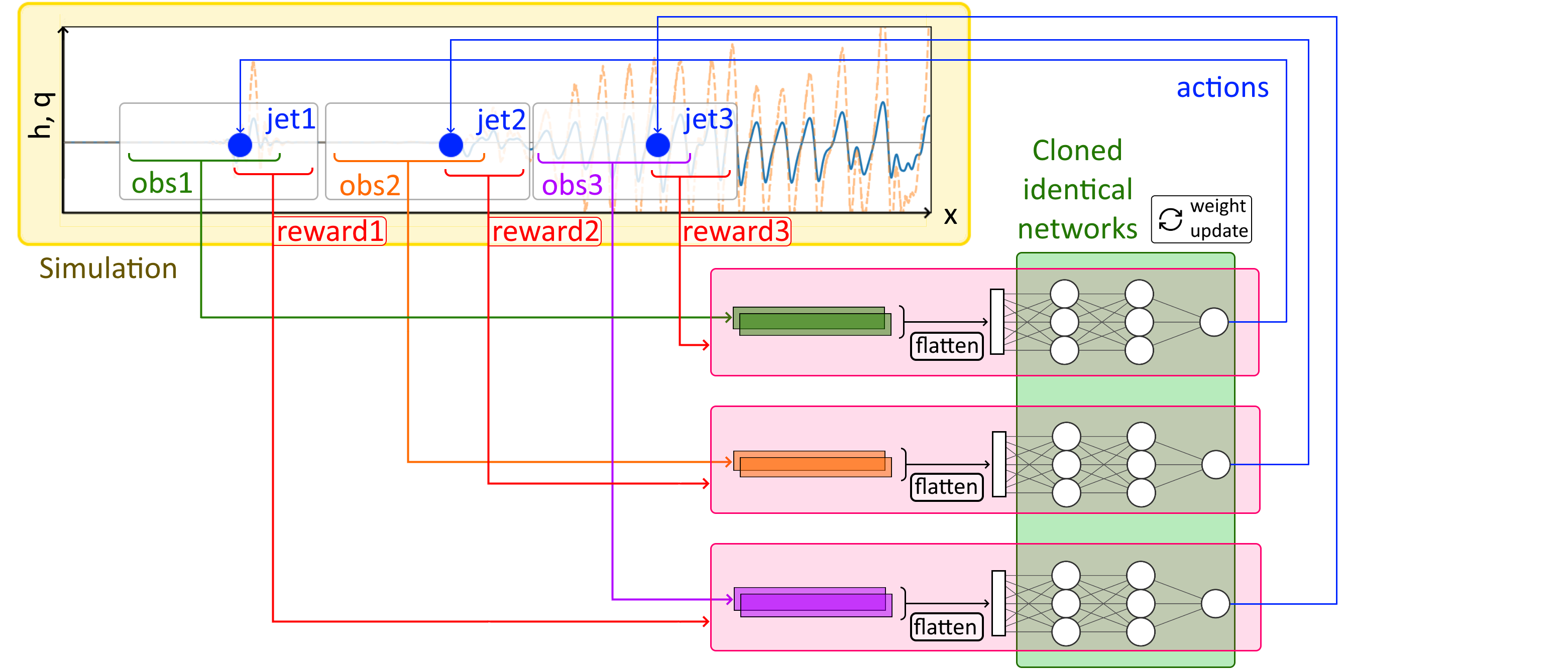}
\caption{\label{fig:all_methods} Illustration of the 3 different methods for control of a system with translational
    invariance and locality. From top to bottom: M1, M2, and M3. M1 is the
    naive implementation of the DRL framework. M2 takes advantage of
    translation invariance of the system to re-use the network coefficients for
    the control of an arbitrary number of jets. M3 both exploits the
    translation invariance, and the locality of the system by using a dense
    reward signal. Details are available in Table \ref{table:methods}.}
\end{center}
\end{figure*}

\begin{table*}[]
\centering
\begin{tabular*}{\textwidth}{@{\extracolsep{\fill}} ||c c c c||} 
 \hline
 Method & states & network & reward \\ [0.5ex] 
 \hline\hline
    \makecell{M1: concatenated jets \\ as a single environment} & \makecell{concatenated \\ and flattened} & MLP & one global reward \\ 
    \hline
    \makecell{M2: convolutional \\ network} & concatenated & \makecell{CNN - equivalent to a \\ MLP on each row of input}  & one global reward \\
    \hline
    \makecell{M3: each jet as a \\ separate environment} & kept separate & \makecell{MLP, shared \\ between jets} & N rewards \\ [1ex] 
 \hline
\end{tabular*}
\caption{Comparison of the 3 different methods for the design of the DRL agent and its interaction
    with the thin liquid film simulation. \\
    MLP = Multi Layer Perceptron, a network where every layer is fully connected. \\
    CNN = Convolutional Neural Network.}
\label{table:methods}
\end{table*}

Machine Learning has become very attractive in the recent years following
several high-impact results of Deep Artificial Neural Networks (DANNs) across
a variety of fields. Results include, for example, attaining super-human
performance at image labeling  \cite{lecun2015deep}, winning against
human professionals at the game of Go \cite{silver2017mastering}, or
achieving control of complex robots \cite{7989385}. Those successes
have demonstrated the ability of DANNs to solve
a wide range of strongly nonlinear, high dimensionality problems that
were resisting investigation using traditional methods. Following these
developments, DANNs are now being applied to other fields of science
including Fluid Dynamics. Recent developments in this domain include,
to name but a few, analyzing laboratory data
\cite{rabault2017performing, cai2019particle}, formulation of reduced order models
\cite{srinivasan2019predictions}, active flow control
\cite{rabault_kuchta_jensen_reglade_cerardi_2019}, the control of
stochastic systems from only partial observations \cite{bucci2019control},
shape optimization \cite{viquerat2019direct}, and closure models for LES and
RANS simulations \cite{beck2018deep}.

More specifically, several of these applications rely on the use of Deep
Reinforcement Learning (DRL). This approach consists in finding, through trial
and error, the solution to a complex problem for which no theoretical or
algorithmic solution is known otherwise. DRL takes advantage of the universal
approximator \cite{hornik1989multilayer} property of DANNs to optimize
interaction with the system it should control through three channels: an
observation of the state of the system, an action taken to control the system,
and a reward function giving feedback on its current performance. This
framework is adapted to cases where only partial, noisy observations of a
stochastic system are available. Therefore, choosing a good reward function is
critical as this is what guides the DANN towards solving a specific problem. In
the following, we will use a specific DRL algorithm known as the Proximal
Policy Optimization (PPO \cite{schulman2017proximal}). This algorithm belongs
to a wider class of algorithms called the Policy Gradient Methods
\cite{sutton2000policy}, and is often regarded as the state-of-the-art
algorithm to be used for control problems where a continuous action space is
present. As the PPO algorithm has been described in details by its initial
authors \cite{schulman2017proximal}, and has been discussed also in the Fluid
Mechanics literature at several occasions
\cite{rabault_kuchta_jensen_reglade_cerardi_2019, garnier2019review}, we refer
the reader curious of more details about the inner working of this algorithm to
these references for further information. Several high-quality implementations
of the PPO algorithm are available open source from public software repositories, and in the
following we will use one of these to provide us will a well-tested
implementation (see Appendix A for further details).

Similarly to \cite{doi:10.1063/1.5116415}, we will in the following use the
word ``action'' to describe the value provided by the ANN based on a state
input, while ``control'' describes the value effectively used in the
simulation. This distinction is especially important as the choice of the
duration of an action, which may extend over several control time steps, is
critical for obtaining good learning (see Figs. 2 and 6 of
\cite{doi:10.1063/1.5116415}). In the following, we use linear interpolation to
determine the value of the control at each time step in-between of action
updates.

In the present work, the system to control is characterized by the high
dimensionality of its output. More specifically, it is natural to use several
jets (up to 20 jets in our simulations, but a larger domain could feature even
more jets). Therefore, using the PPO algorithm effectively becomes challenging.
Indeed, the naive approach which consists in using a single network with
several outputs does not scale well to an increasing number of jets, as the
combined combinatorial size of the output domain for N jets grows as a power of
N, and therefore the curse of dimensionality is a threat to finding effective
control strategies. However, one can observe that the system to control
features a translation invariance along the $x$-axis. Therefore, one should be
able to take advantage of this property to optimize learning, in the same way
that Convolutional Neural Networks (CNNs) take advantage of translational
invariance of 2D images across the $x$- and $y$-directions to share
convolutional kernels across the whole image and therefore reduce the number of
weights needed and improve learning performance \cite{lecun1995convolutional,
krizhevsky2012imagenet}.

Following this observation, we design three different methods for performing
control of the system:

\begin{itemize}
    \item First, a `naive' method in which the input regions from all jets are
        concatenated and flattened before being provided to the network, and the
        dimensionality of the output is equal to the number of jets. In this
        case, the reward is evaluated over the whole combined reward region.
        This method will be referred to as the Method 1 (`M1') in the following.

    \item Second, we apply control following a method that is a direct analogy
        of the CNN used in image analysis. In this case, the inputs from the
        regions around different jets are concatenated without flattening, and
        fed into a purely convolutional network. This allows to apply the exact
        same weights, and therefore the same policy, on all inputs to generate
        the individual jet values. There also, only one global reward is
        available, similar to M1. Due to purely technical implementations
        difficulties and the exact architecture of the DRL framework, this is however not
        implemented as a CNN in practise, but as a formally equivalent cloned
        network. This method will be referred to as the Method 2 (`M2').

    \item Third, we apply control by splitting the simulation into several DRL
        environments, i.e we consider each triplet [jet observation domain, jet
        value, jet reward domain] as a separate environment. A unique agent is
        sampling trajectories from these environments as if they were clones of
        the same environment, taking advantage of the translational invariance
        of the system. Similarly to M2, the same policy is applied on all the
        jets. However, in contrast to both M1 and M2, this method effectively
        `densifies' the reward: instead of performing learning based on 1 single
        global reward, many individual rewards are obtained (one for each jet),
        providing more granularity in the learning process. This will be
        referred to as the method 3 (`M3').
\end{itemize}

Those 3 different methods for controlling several jets are summarized in table
\ref{table:methods}, and presented in Fig. \ref{fig:all_methods}. Note that in
all cases the architecture of the network is kept equivalent (except for the
output layer in case M1 vs. M2 and M3), and only the translational invariance
and reward densification differ between those methods.

As visible in Table \ref{table:methods} and Fig. \ref{fig:all_methods}, the
methods M1, M2, and M3 reflect increasingly the structure of the underlying
system to control, and therefore we expect in terms of learning speed and
performance that M1 < M2 < M3, where the order relation describes `how good'
and `how fast' the policies and trainings are. This hypothesis is confirmed
experimentally in the next section.

\section{Results}

\subsection{Physical metaparameters and successful learning}

Using the methodology presented in the previous sections, together with a
consistent set of metaparameters, satisfactory learning is obtained. We find
that tuning the metaparameters of the PPO algorithms is not crucial to obtain
learning, and in all the following we will use the default PPO metapameters
recommended by the package used. This is in good agreement with other studies,
that have generally observed that the PPO algorithm is quite robust to the
exact value of its metaparameters. By contrast, the `physical' metaparameters
of the simulation setup are important. In the following, the parameters used
(unless stated otherwise) correspond to a duration of action $\Delta t_{action}
= 0.05$, i.e. 50 steps of the numerical solver are performed between each
action update, which corresponds to a typical propagation of the waves by a
distance of the order of $\Delta x = 0.2$. This is typically 10 percents of the
half-width of a jet, $w_i = 2$, which itself is typically around 10 percent of
the wavelength of big fully developed waves $\lambda = 20$. The duration of an
episode, which dictates the number of actions performed between learnings, is
set to $\Delta t_{episode} = 20$. This allows to sample trajectories in the
phase space that are long enough that the effect of policy updates can be
observed. Finally, the reference maximum strength of a jet is set to be $M =
5$. While the exact numerical relation between these quantities is not
critical, their relative orders of magnitudes must be respected to be able to
control the system. For example, using too wide jets (larger than the typical
size of the waves) obviously does not allow to perform control. Similarly, too
small jets are not enough to significantly alter the propagating waves. The
choice of $\Delta_t$ is also critical to allow the discovery of a valid policy
through trial and error, similarly to what has been discussed in for example
\cite{doi:10.1063/1.5116415} and is illustrated later in this section.

\begin{figure*}
    \begin{center}
        \includegraphics[width=0.79\textwidth]{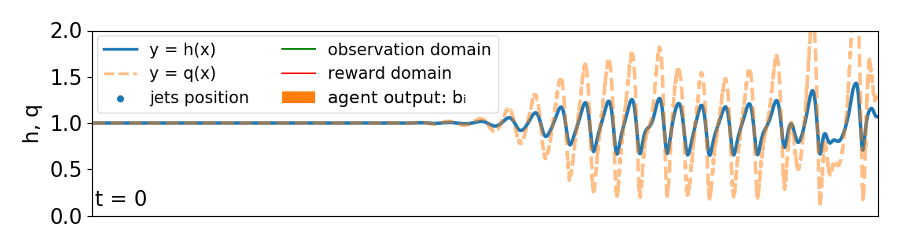} 
        \includegraphics[width=0.79\textwidth]{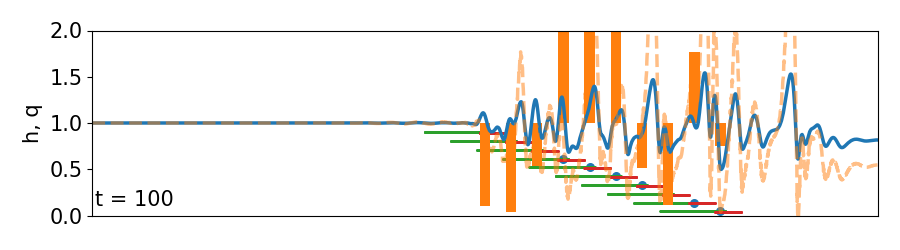} 
        \includegraphics[width=0.79\textwidth]{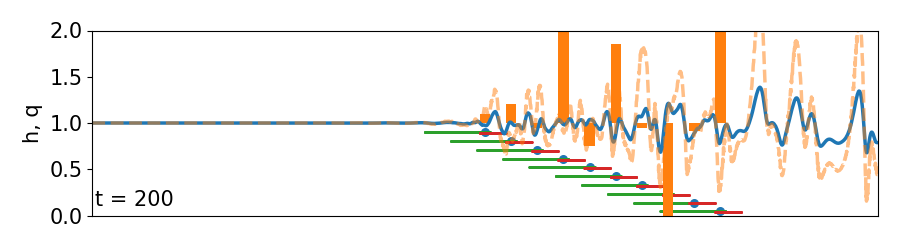} 
        \includegraphics[width=0.79\textwidth]{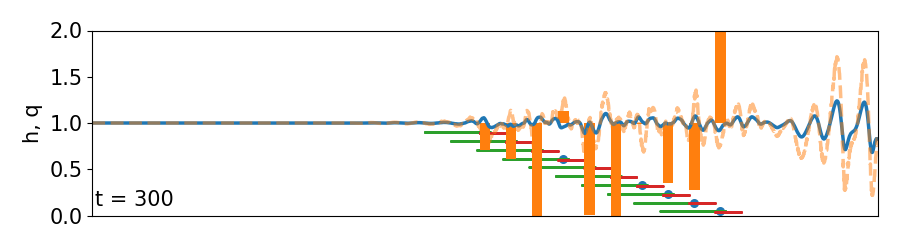} 
        \includegraphics[width=0.79\textwidth]{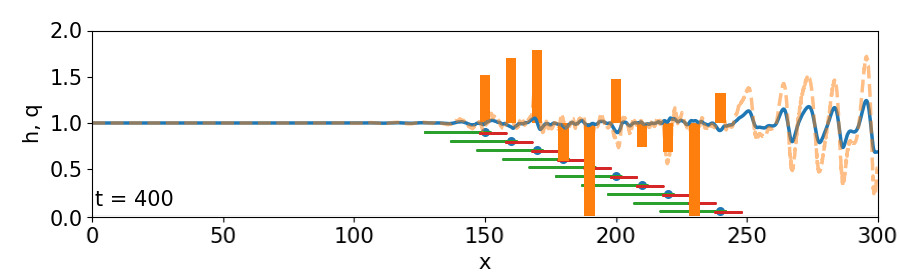} 
        \caption{\label{fig:good_training_1} Evolution of the simulation
        during the training phase. Here we are using the method M3 with 10
        jets, coupled to one single simulation, and default physical
        metaparameters (see discussion in the text). We can see that an
        efficient policy has been found already for a non-dimensional time of
        around $t=400$. This typically takes less than three minutes on a
        recent CPU, using a single core. 
       }
    \end{center}
\end{figure*}

In this section, we only present results obtained with the training method M3,
which is the best performing one (see discussion in section III.B). Successful
learning, corresponding to the default parameters, is illustrated in Fig.
\ref{fig:good_training_1}. There, the method M3 is used to train 10 jets to
perform active control of the incoming waves. As visible in Fig.
\ref{fig:good_training_1}, the ANN can effectively kill waves on the control
region.

\begin{figure*}
    \begin{center}
        \includegraphics[width=0.95\textwidth]{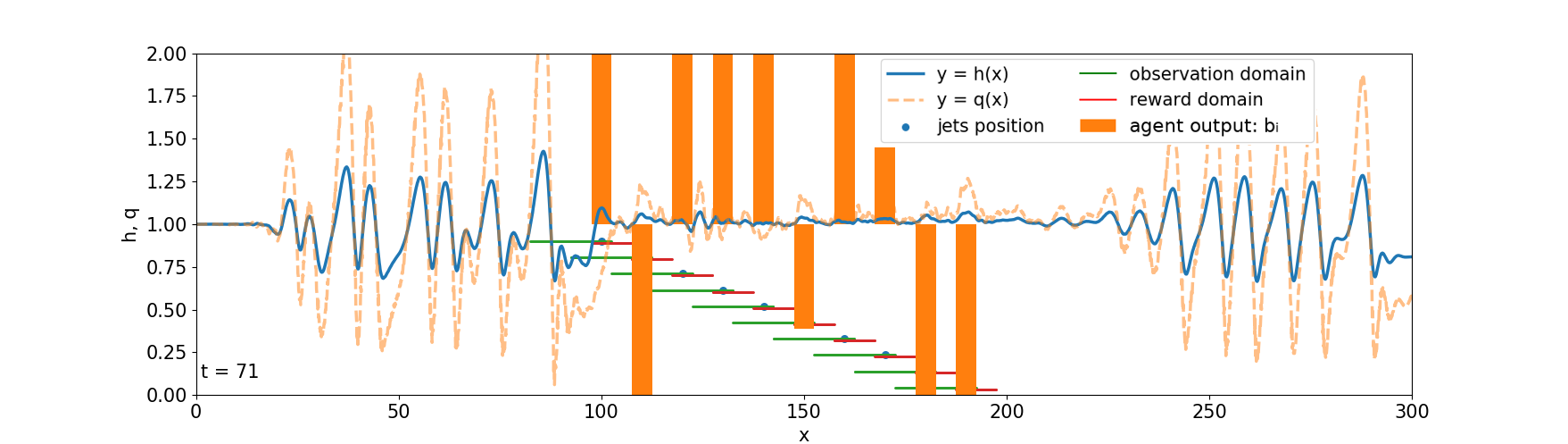}
        \includegraphics[width=0.95\textwidth]{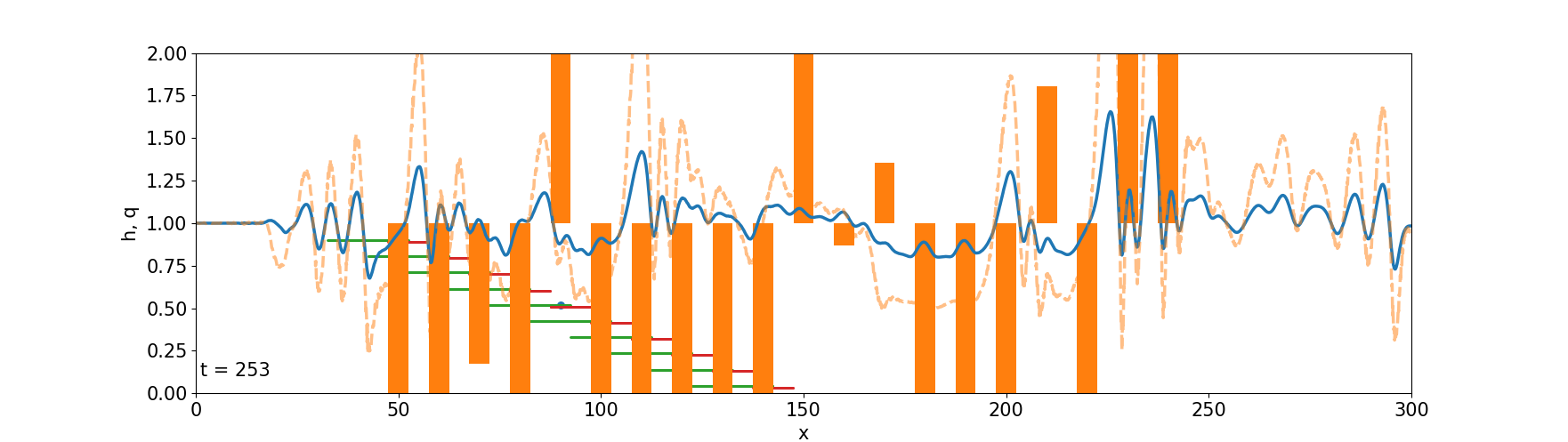}
        \caption{\label{fig:good_training_2} Illustration that control can be successfully applied
        even when chaotic, fully developped waves are used as an input to the
        control region, as long as the physical metaparameters used are
        relevant. Top: render of a policy trained over 150 episodes, acting
        while a stochastic perturbation jet is present at $x=20$, creating a
        chaotic region. There we are using the method M3 with 10 jets, and the
        standard jet strength. We can see that large incoming waves are
        effectively controlled. Bottom: render of a policy trained during 740
        episodes, using the same method and with the same perturbation jet at
        $x=20$. We use twice as many jets as in the previous trainings (top),
        but each jet is 10 times less powerful (reduction of $M$ by a factor of
        10). We observe that the policy fails to fully dissipate the large
        waves (some level of control is still achieved, though), as the control
        strength is not sufficient to compensate for the wave growth.}
    \end{center}
\end{figure*}

As visible on Fig. \ref{fig:good_training_1}, the placement of the jets in the
physical domain is there such that, upon control of the system, the waves never
get the possibility to fully develop into a chaotic regime. This means that,
upon successful control, the problem becomes even simpler for future actuation
as only small waves are present, which are less nonlinear than large chaotic
waves. To test the ability of the system to learn and control also large,
chaotic waves, we run trainings with a strong perturbation jet added at $x=20$.
The perturbation jet is sampled from a uniform distribution on the range [-5;
5]. Typical results are visible in Fig. \ref{fig:good_training_2}. One can see
that, in this case, satisfactory control can also be obtained (Fig.
\ref{fig:good_training_2}, top). However, this holds only if the jets are made
strong enough (no satisfactory control is obtained for Fig.
\ref{fig:good_training_2} bottom), while with the configuration of Fig.
\ref{fig:good_training_1} even weak jets were enough to exert effective control
(see next paragraph).

\begin{figure*}
    \begin{center}
        \includegraphics[width=0.45\textwidth]{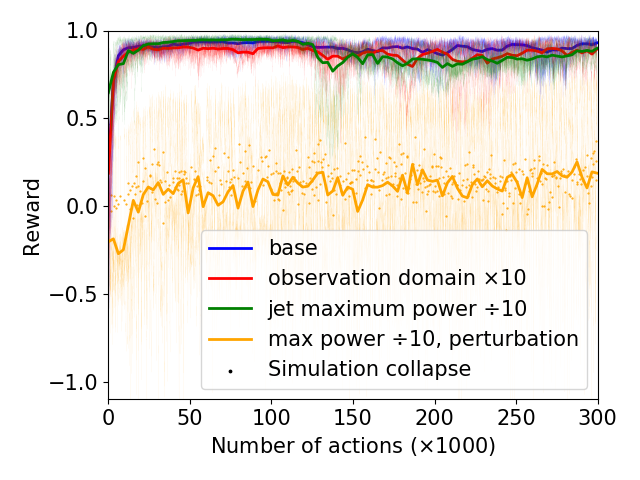}
        \includegraphics[width=0.45\textwidth]{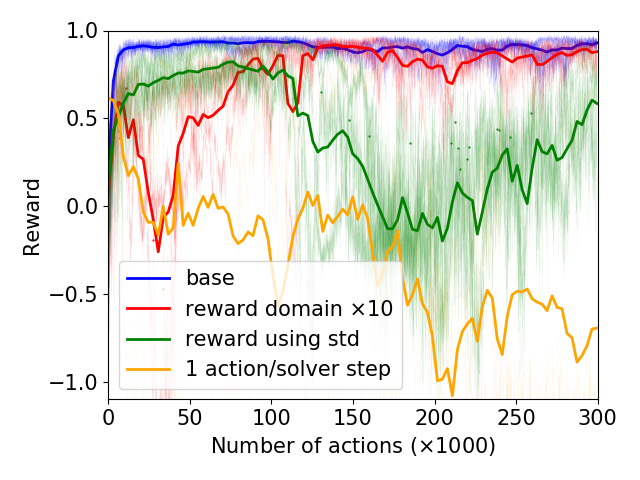}
        \caption{\label{fig:effect_phys_metaparameters} Analysis of the effect of both physical and DRL
        metaparameters on the strategies found. The baseline configuration
        corresponds to 4 evenly-spaced jets, $w_i = 2.5$, $L_{obs} = 10$,
        $L_{reward} = 10$, and the agent being trained with method M3. The
        first jet is located at $x=150$, and the spacing between the jets is
        $10$. The reward on vertical axis is computed with the same function on the
        same reward domain for all the trainings. There is no perturbation jet
        in the base case. The `simulation collapse' label corresponds to points
        where bad choice of jet strength by the ANN leads to numerical
        instability of the simulation, in which case the simulation is
        resetted. Each thick learning curve is the average of 3 trainings
        (individually shown as thin colored curves). On the left, we
        investigate the effect of the observation domain size and jet strength
        on the learning quality. We observe that the size of the observation
        domain has little effect on the learning, as the ANN is able to select
        the relevant information. Similarly, in the case with no perturbation
        jet, the waves are small enough that the strength of the jets can be
        reduced and control is still obtained. By contrast, if a perturbation
        jet is used, the waves are too big to be controlled with the weakest
        jets. On the right, we investigate the effect of the reward parameters
        and the number of solver steps per action. We observed that using a
        reward domain that is too large, i.e. includes a large region that is
        too far away from the jets to be initially controlled, disturbs the
        learning and that more time is needed in this case to find a good
        policy. Similarly, a reward function based on using a standard
        deviation works less well, as the ANN can try to change the mean level
        of the flow. Finally, using a duration for actions that is far smaller
        than the natural period of the system (1 numerical timestep per action)
        completely stops the learning, similarly to what had been observed in
        \cite{doi:10.1063/1.5116415}.}
    \end{center}
\end{figure*}

The effect of more metaparameter experimentations are presented in Fig.
\ref{fig:effect_phys_metaparameters}. There, we present learning curves based
on evaluation from the reward function Eqn. (\ref{eqn:reward}), even in the
case when another reward function is used during training. As visible in Fig.
\ref{fig:effect_phys_metaparameters}, the exact size of the observation domain
for each jet is not critical for the learning. This is consistent with previous
reports that DRL is usually good at filtering out unnecessary information.
Similarly, in the default configuration, the maximum strength of the jets is
not too critical. This has already been discussed, and corresponds to the fact
that upon discovery of a successful strategy by the ANN, waves can be killed
before they fully develop - therefore, requiring only weak jets for successful
control. However, as was illustrated in Fig. \ref{fig:good_training_2}, this is
not the case if the incoming waves are strong enough. By contrast, the choice
of the reward domain, reward function, and action update frequency are much more
important to obtain successful and efficient training, as illustrated by the
second plot of Fig. \ref{fig:effect_phys_metaparameters}. This is, there also,
consistent with previous reports, such as
\cite{rabault_kuchta_jensen_reglade_cerardi_2019, doi:10.1063/1.5116415}, and
can be easily understood in each of the cases presented. Indeed, using a too
large reward domain means that, until the waves are successfully killed on a
large region, a lot of the reward signal is uncorrelated with the individual
action of each jet - as it incorporates many waves from far downstream each
individual jet. Similarly, using the standard deviation of the water height $\text{std}(h)$, instead of the
deviation to the reference water $\sum\limits_{x \in A_{reward}} [\mathrm{h}(x,
t)-1]^2$ in Eqn. (\ref{eqn:reward}), means that the agent may try to reduce the
waves fluctuations around a different mean water level as forcing locally
changes the mean value of the water height. Therefore, this confuses the agent during
learning. Finally, the most drastic effect on learning is observed when the action
period is reduced to be equal to the numerical timestep. Similarly to
\cite{rabault_kuchta_jensen_reglade_cerardi_2019, doi:10.1063/1.5116415}, this
means that only white noise forcing is applied in general to the system, which
fails at finding any consistent strategy.

\subsection{Comparison of the three training methods M1, M2, and M3}

\begin{figure*}
\begin{center}
\includegraphics[width=.45\textwidth]{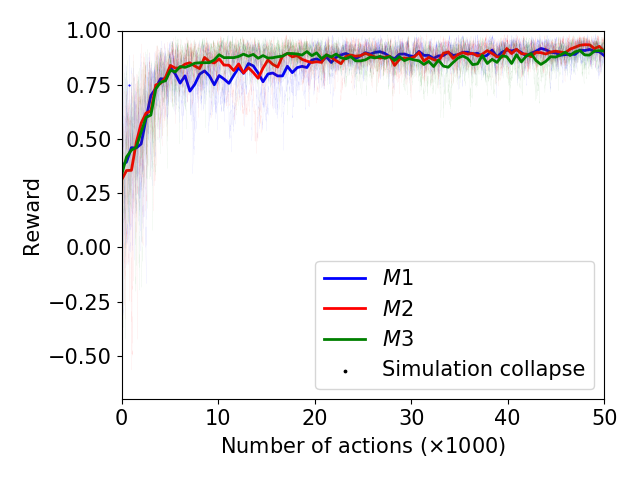}
\includegraphics[width=.45\textwidth]{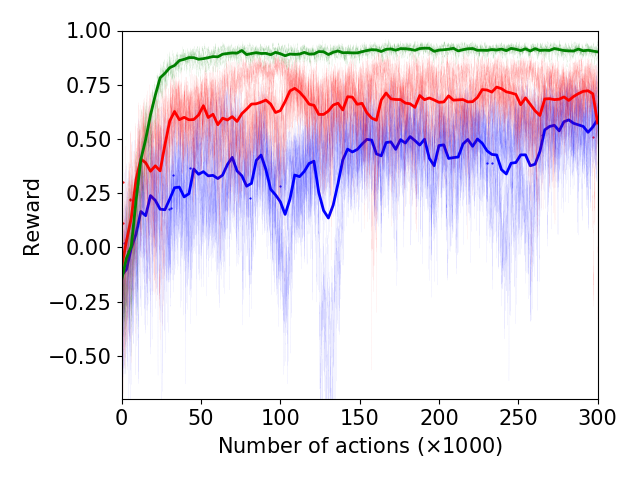}
\includegraphics[width=.45\textwidth]{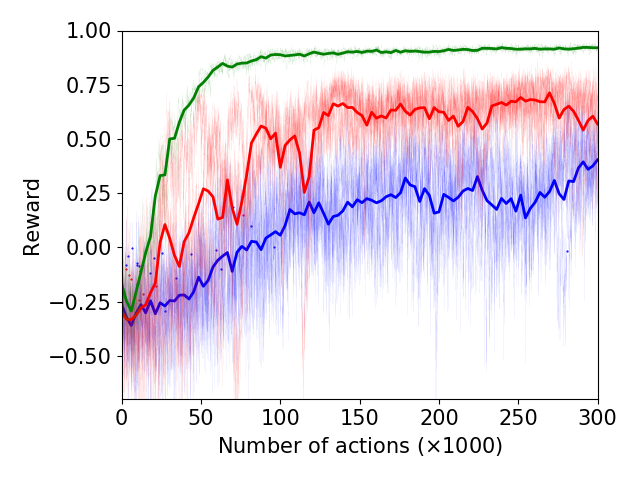}
\includegraphics[width=.45\textwidth]{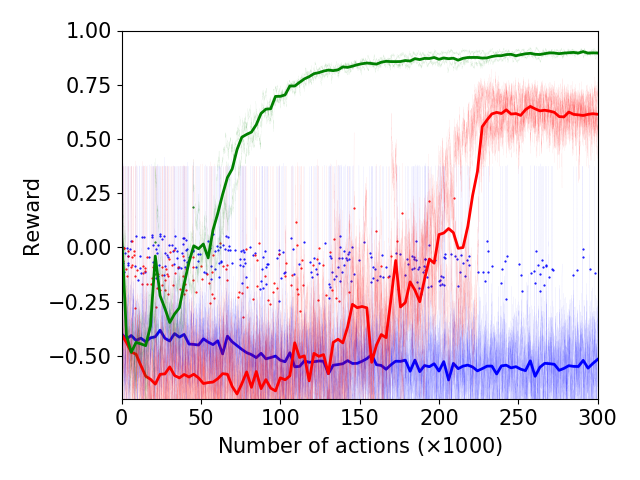}
    \caption{\label{fig:results_all_methods} Comparison of the efficiency of the learning (both speed and
    quality of the final policy) for the methods M1, M2, M3 (varying color) and
    an increasing number of jets from left to right, top to bottom
    (respectively 1, 5, 10 and 20 jets). The 'simulation collapse' label
    corresponds to points where bad choice of jet strength by the ANN leads to
    numerical instability of the simulation, in which case the simulation is
    resetted. Each thick learning curve is the average of 3 trainings
    (individually shown as thin colored curves). As visible here, the method M3
    is best, with increasing advantage over both M2 (second best) and M1
    (worst) as the number of jets increases.}
\end{center}
\end{figure*}

\begin{figure*}
\begin{center}
\includegraphics[width=.45\textwidth]{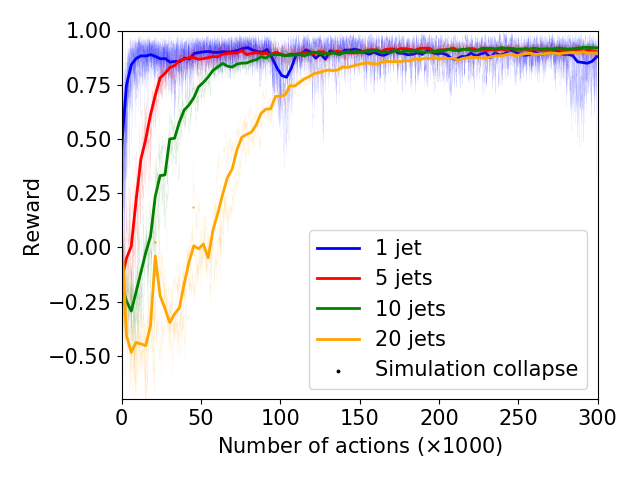}
\includegraphics[width=.45\textwidth]{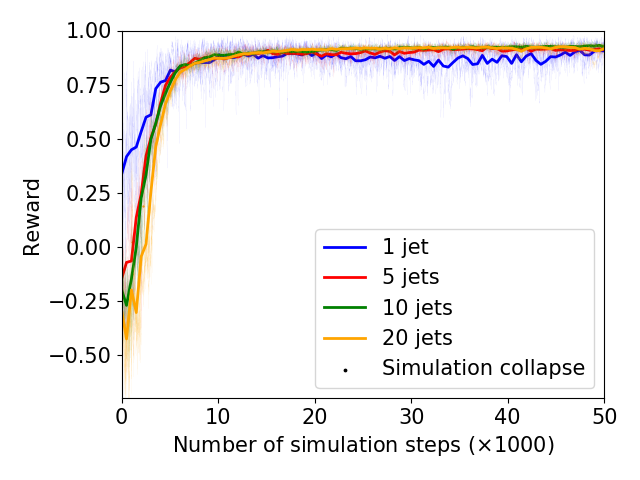}
\caption{\label{fig:focus3} Detailed overview of the training of the DRL agent following the method
    M3. On the left, the same data as in Fig. \ref{fig:results_all_methods} are
    reproduced, showing that more actions are needed to perform satisfactory
    learning with increasing number of jets. However, we show on the right that
    the learning as a function of the number of simulation steps, i.e.
    advancement in time of the simulation, remains equivalent between all
    cases. This is because, with method M3, the number of actions per
    advancement in time of the simulation is proportional to the number of
    jets, i.e. `number of actions $\propto$ N $\times$ number of steps', and
    therefore using more jets allows to extract more individual triplets of
    (state, action, reward) out of each advancement in time of the simulation.
    More specifically, we observe that learning with the method M3 uses
    constant CPU resources (which is proportional to the number of steps rather
    than the number of actions), independently of the number of jets used.}
\end{center}
\end{figure*}

Learning curves for a varying number of jets (1, 5, 10, and 20 jets) and the
three different methods are presented in Fig. \ref{fig:results_all_methods}. In
addition, since several actions are obtained for each numerical advancement of
the simulation in the case M3, the data in this later case are presented again
in Fig. \ref{fig:focus3}, but showing on the horizontal axis both the number of
actions and the number of numerical advancements. It is visible there also that
the DRL agent is able to apply effective control on the system. The evolution
of the reward during training indicates that several phases take place. As
should be expected, control with a random policy (as takes place at the
beginning of each training) degrades the reward compared with the case without
control (the reward without control is around 0, and in the first phase of
training a reward as low as -0.5 can be observed, corresponding to larger waves
being obtained when bad control is applied). However, as training takes place,
the reward starts to increase at least in the cases when training is
successful. Finally, a plateau in performance is reached upon successful
training (or failure of training). The value of the reward, that is close to 1
in several cases, indicates that the system is controllable, and that this
control is close to perfect in the sense that it manages to kill close to all
fluctuations in $h$ (see Eqn. (\ref{eqn:reward})), i.e. all waves are canceled.

However, it is clear that there are large variations between the efficiency of
the different methods. While all methods perform similarly in the case with one
single jet, which is really a consistency test for the 3 methods as they are
all equivalent in this particular case, differences appear when the number of
jets starts to be increased. More specifically, one can observe that as the
number of jets increases, methods M1 and M2 see a reduction of their efficiency
regarding both the speed of convergence, and the quality of the control
strategy asymptotically found. It appears that the method M1 is performing
worst, while method M2 is doing slightly better though also degrading with
increasing number of jets. By contrast, method M3 sees close to no reduction in
performance when increasing the number of jets (at least normalizing by the
number of simulation advances as shown in Fig. \ref{fig:focus3}, which is
proportional to the CPU cost, instead of the number of actions taken).
Generally, this confirms experimentally that here we clearly observe that $M1 <
M2 < M3$, where the ordering relation describes effectiveness of the methods.

The difference in efficiency between these 3 methods can easily be understood,
as hinted in the previous section, by considering both the invariance of the
system by translation and how this compares to the architecture of the DRL
agent, as well as the amount of fine-grained reward signal received.

Firstly, the methods 1 does not reflect whatsoever the invariance by
translation of the physical system. While this has no consequences in the case
when only 1 jet is used, this drastically reduces the ability of the agent to
learn when more jets are present. Indeed, this means that the network has to
`learn from scratch' by trial and error the policy applied to each jet, and
there is no sharing of the weights of policies found at different locations.
Therefore, the method 1 is subject to the curse of dimensionality. If, for the
sake of a thought experiment, one considers that the action space for each jet
is a discrete set of $p$ values in the admissible range, then the method 1 may
need typically up to $C \times p^N$ trials to sample effectively the policy in
the case where $N$ jets are used, where $C$ is a constant. By contrast, methods
2 and 3 use the exact same set of weights to link the state and jet control at
each position, either it is by using a fully convolutional network or a shared
agent, and therefore they escape this curse of dimensionality.

Secondly, both methods 1 and 2 fail to take into account that the system
presents some locality that allows, if exploited correctly, to `densify' the
reward. By contrast, the method M3 takes into account this locality, and is
therefore able to extract N reward signals instead of 1, therefore collecting
much more information driving the gradient descent. What is meant here is that,
while the output flow conditions obtained after the jet number $j$ do influence
what happens at the area around the jet number $j+1$, the actuation has first
and foremost a short term effect on the flow around the position where it is
applied. Therefore, it does make sense to consider the neighborhood of each jet
independently, and use it in an individual DRL control loop. The approach
chosen in M3, which consists in having an agent learn from the observation and
reward of each jet, takes therefore full advantage of both the invariance and
locality properties of the system. As visible in Fig. \ref{fig:focus3} this
means that, while more actions are needed to learn a valid policy as the number
of jets N is increased using the method M3, since at the same time the number
of actions executed by numerical advancement of the simulation is N, the
learning takes place in constant number of numerical advancements, i.e.
constant CPU time when the simulation is the leading computational cost (which
is usually the case in Fluid Mechanics, see the discussion in
\cite{doi:10.1063/1.5116415}). By contrast, the methods M1 and M2 are at a
double disadvantage: first, they receive less volume of reward, which is the
signal allowing to perform training, i.e. less information is fed into the DRL
algorithm. Second, the reward in the cases M1 and M2 covers a very large area,
which encompasses several jets, and therefore the feedback information gets
less representative of the actual state of the system. Indeed, if one jet
performs a `good' action and another a `bad' one at the same time, as a result
the reward will be average, and the DRL algorithm has no way to know that it
actually performed well on one jet, and poorly on the other.

However, one may argue that the densification of the reward used in method M3
may also be a potential problem for the optimality of the solution found.
Indeed, it means that all rewards are obtained on a local, rather than a
global, basis. In our case, this is not a problem, as the optimal strategy at a
local level is also the optimal strategy at a global one. However, this may be
a problem for method M3, if it is used exactly as deployed here, in a more
complex system where the local and global optimization processes are in
conflict with each other. One could, however, easily mitigate such an issue, by
defining each local reward as a weighted average of the true local reward, and
the global reward taken over the whole system.

\section{Conclusion}

We present the first successful control of the falling liquid film flow through
a 1D simulation, using DRL. In addition to proving that the system is
controllable, we show that the DRL methodology can be used in such a way that
it handles an arbitrary number of jets. Therefore, one can effectively escape
the curse of dimensionality on the control output size. This relies on
satisfactorily exploiting invariant and locality properties of the underlying
system. Failing to exploit one, or several, of these properties leads to
reduced quality of the learning and of the final policy. While this is the
first time, to our knowledge, that this methodology is proposed for the optimal
control of physical systems, it is deeply inspired by the success of CNNs in
image analysis. Indeed, CNNs prove efficient in such tasks by similarly taking
advantage of translation invariance of image semantic content.

This work, possibly combined together with the results previously obtained in
\cite{doi:10.1063/1.5116415}, opens the way to applying DRL to more realistic,
complex physical systems. Indeed, such systems may require many control outputs
to be manipulated, which is a difficulty in itself due to the curse of
dimensionality. However, those same systems usually present many properties of
locality (either strong or weak) and invariance, therefore the kind of
techniques presented here can be envisioned as a solution to this
dimensionality problem. We expect that such trainings, that will resort on the
use of both several independent simulations in parallel similar to
\cite{doi:10.1063/1.5116415}, and environment splitting and / or convolutional
policy as presented in the present work, may be able to scale to several
thousands of CPUs during training and become a tool for the study of realistic
flow configurations.

\section*{Acknowledgements}

We gratefully acknowledge discussions with Dr. Bin Hu, who put Dr. Jean Rabault and
Pr. Zhizhao Che in contact a few months before this project was initiated. Discussions
and support received from Pr. Atle Jensen are gratefully acknowledged.
The help of Terje Kvernes for setting up the computational infrastructure used
in this work is gratefully acknowledged. We gratefully acknowledge help,
guidance, and CPU time provided by UH-IaaS (the Norwegian Academic Community
Cloud, IT departments, University of Oslo and University of Bergen;
http://www.uh-iaas.no/) when setting up the computational infrastructure. This
work was performed thanks to funding received by the University of Oslo in the
context of the `DOFI' project (grant number 280625).

\section*{Appendix A: Open Source code release}

The source code of this project, together with a docker container that enforces
full reproducibility of our results, is released as open-source on GitHub [NOTE:
the repository is empty for now, the code will be released upon publication in
the peer-reviewed literature]:
\textit{https://github.com/vbelus/falling-liquid-film-drl}. The PPO agent is
based on the open-source implementation provided by stable-baselines \cite{stable-baselines},
which builds on top of the Tensorflow framework \cite{abadi2016tensorflow}.
We are using the RL toolkit OpenAI Gym to build custom environments and interact
with the agent \cite{1606.01540}.

% \section{bibliography}
\bibliographystyle{unsrt}
\bibliography{BibliographyBib}

\end{document}